\documentclass[12pt]{article}
\usepackage[latin9]{inputenc}
\usepackage{amsmath,amsfonts,amssymb,amsthm,amstext,amscd,array}
\usepackage[table]{xcolor}
\usepackage{tikz}
\usepackage{mathrsfs}
\usepackage{bbm}
\usepackage{bm}
\usepackage{indentfirst}
\usepackage[arrow,matrix,curve]{xy}
\usepackage{hyperref}
\usepackage{refstyle}
\usepackage{cite}

\renewcommand{\thefootnote}{\fnsymbol{footnote}}
\newcommand{\newsection}{\setcounter{equation}{0}\section}

\makeatother

\begin{document}
\begin{titlepage}
\vspace{3cm}
\baselineskip=24pt

\begin{center}
\textbf{\LARGE{} Semi-simple enlargement of the $\mathfrak{bms}_3$ algebra from a {\Large $\mathfrak{so}(2,2)\oplus\mathfrak{so}(2,1)$} Chern-Simons theory}
\par\end{center}{\LARGE \par}

\begin{center}
\vspace{1cm}
 \textbf{Patrick Concha}$^{\ast}$,
\textbf{Nelson Merino}$^{\dag}$,
\textbf{Evelyn Rodríguez}$^{\ddag}$,
\\[3mm]\textbf{Patricio Salgado-Rebolledo}$^{\ast}$
\textbf{and} \textbf{Omar Valdivia}$^{\sharp}$
\footnotesize
	\\[5mm]
	$^{\sharp }$\textit{Facultad de Ingeniería y Arquitectura,}\\
	\textit{ Universidad Arturo Prat, Iquique-Chile.}
	\\
	$^{\ast}$\textit{Instituto
		de Física, Pontificia Universidad Católica de Valparaíso, }\\
	\textit{ Casilla 4059, Valparaiso-Chile.}
	\\
	$^{\dag}$\textit{Univ Lyon, ENS de Lyon, Univ Claude  Bernard, CNRS, Laboratoire de Physique,} \\\textit{F-69342 Lyon, France}\\
	$^{\ddag}$\textit{Departamento de Ciencias, Facultad de Artes Liberales,} \\
	\textit{Universidad Adolfo Ibáñez, Viña del Mar-Chile.} \\[5mm]
	\footnotesize
	\texttt{patrick.concha@pucv.cl},
	\texttt{nelson.merino@ens-lyon.fr},
	\texttt{evelyn.rodriguez@edu.uai.cl},
	\texttt{patricio.salgado@pucv.cl},
	\texttt{ovaldivi@unap.cl}
	\par\end{center}
\vskip 15pt
\begin{abstract}
In this work we present a BMS-like ansatz for a Chern-Simons theory based on the semi-simple enlargement of the Poincaré symmetry, also known as AdS-Lorentz algebra. We start by showing that this ansatz is general enough to contain all the relevant stationary solutions of this theory and provides with suitable boundary conditions for the corresponding gauge connection. We find an explicit realization of the asymptotic symmetry at null infinity, which defines a semi-simple enlargement of the $\mathfrak{bms}_3$ algebra and turns out to be isomorphic to three copies of the Virasoro algebra. The flat limit of the theory is discussed at the level of the action, field equations, solutions and asymptotic symmetry.

\end{abstract}
\end{titlepage} \setcounter{page}{2}

\newpage{}

{\baselineskip=12pt \tableofcontents{}}

\global\long\def\thefootnote{\arabic{footnote}}
 \setcounter{footnote}{0}

\newsection{Introduction}
Different Chern-Simons (CS) (super)gravity models based on extensions and deformations of the Poincaré and AdS algebras have been recently introduced in the literature~\cite{Edelstein:2006se,Izaurieta:2006aj,Izaurieta:2009hz,Gonzalez:2014tta,Salgado:2014jka,Concha:2014vka,Concha:2014zsa,Fierro:2014lka,Concha:2015woa,Aviles:2018jzw}. In general, apart from the vielbein and spin connection, these theories include new gauge fields which appear as a direct consequence of the enlargement of the symmetry. Although such models represent interesting gravity theories that extend General Relativity (GR), not much is known yet about their solutions and their physical interpretations. In three-dimensions, some progress has been recently carried out for the so-called Maxwell algebra~\cite{Schrader:1972zd,Bacry1970,Gomis:2017cmt} and its semisimple counterpart~\cite{Soroka:2011tc,Gomis:2009dm}, also known as AdS-Lorentz algebra. The general solution in the stationary sector was reported in~\cite{Hoseinzadeh:2014bla} and, as a direct consequence of a symmetry enhancement, it depends on three arbitrary functions for which no gauge fixing was discussed. The study of non-stationary solutions with null boundary in a CS gravity theory with Maxwell symmetry, was first reported in~\cite{Concha:2018zeb}. The general solution was found by means of a suitable ansatz for the gauge connection consisting in the standard Bondi-Metzner-Sachs-van der Burg (BMS) gauge for the space-time metric \cite{PhysRev.128.2851} plus a suitable choice for the extra field content.

Asymptotic symmetries in gravitational theories, on the other hand, have attracted great attention in the last decades due to their relation to different aspects of quantum gravity \cite{Strominger:1997eq,Banados:1998gg,Arcioni:2003xx,Barnich:2010eb,Barnich:2012xq,Strominger:2013jfa,Ashtekar:2014zsa,Donnay:2015iia,Oblak:2016eij}. Along these lines, three-dimensional gravity is of particular interest. Despite the absence of local degrees of freedom, CS gravities in three dimensions present a rich boundary dynamics and provide toy models that could realize the bulk/boundary duality beyond the AdS/CFT scenario~\cite{Maldacena:1997re}. In fact, in the case of three-dimensional asymptotically AdS Einstein gravity, it is possible to define an infinite number of conserved charges at spatial infinity, which span a central extension of the two-dimensional conformal algebra \cite{Brown:1986nw}. In the asymptotically flat case, an infinite dimensional asymptotic symmetry algebra can be found at null infinity, given by the $\mathfrak{bms}_3$ algebra \cite{Barnich:2006av}. These infinite-dimensional algebras define the global symmetries of the corresponding boundary field theory that controls the solution space of the bulk theory \cite{Coussaert:1995zp,Barnich:2013yka} (for a geometric point of view see also \cite{Barnich:2017jgw}). Different generalizations of asymptotic symmetries for supersymmetric and higher spin extensions of three-dimensional Einstein gravity, have been found in the last years \cite{Henneaux:1999ib,Troessaert:2013fma,Gonzalez:2014tba,Barnich:2014cwa,Fuentealba:2015wza,Banerjee:2016nio,Detournay:2016sfv,Setare:2017mry,Fuentealba:2017fck,Banerjee:2018hbl}. In \cite{Concha:2018zeb}, the surface charge algebra of a CS gravity theory with Maxwell symmetry was presented. In this case, the asymptotic symmetry is given by an enlargement of the $\mathfrak{bms}_3$ algebra.

In this paper, we extend these results to the case of a three-dimensional gravity theory invariant under the AdS-Lorentz algebra. We construct a BMS-like gauge for the field content of the theory, and find that the asymptotic symmetry of conserved charges at null infinity turns out to be given by a semi-simple enlargement of $\mathfrak{bms}_3$. This infinite-dimensional symmetry was recently introduced as an expansion of the Virasoro algebra in~\cite{Caroca:2017onr}.
These considerations also indicate how to gauge fix the arbitrary functions of the stationary solution found in~\cite{Hoseinzadeh:2014bla} in such a way that it is contained in the solution space defined by the boundary conditions proposed here. As a consistency test, we also show that a flat limit of the asymptotic symmetry recovers the results previously found in~\cite{Concha:2018zeb}.

This paper is organized as follows. In Section \ref{adslg}, we review the main properties of a CS theory invariant under the AdS-Lorentz algebra. In Section \ref{solutions}, we propose a special gauge fixing for the known stationary solutions of the theory, and we propose a BMS-like gauge for the full field content that allows one to solve the field equations. We show that the stationary solutions previously discussed are contained in the solution space with BMS-like boundary conditions. In Section \ref{asym}, we use the boundary conditions to compute the conserved charges of the theory and the corresponding asymptotic symmetry algebra. Some aspects about the flat limit are also discussed. Finally, in Section \ref{comments} we conclude with a discussion of the results and possible future directions.

\newsection{Three-dimensional AdS-Lorentz gravity}
\label{adslg}

\subsection{The AdS-Lorentz algebra}

An interesting semi-simple enlargement of the Poincaré symmetry is the AdS-Lorentz algebra \cite{Soroka:2006aj,Salgado:2014qqa}, which can be obtained by introducing
an additional set of generators $Z_{a}$ that render the translations in the Poincaré algebra non-abelian. The commutation relations of
this algebra read
\begin{equation}
\begin{tabular}{lll}
 \ensuremath{\left[J_{a},J_{b}\right]=\epsilon_{abc}J^{c}\,,}  &  \medskip &  \ensuremath{\left[P_{a},P_{b}\right]=\epsilon_{abc}Z^{c}\,},\\
 \ensuremath{\left[J_{a},Z_{b}\right]=\epsilon_{abc}Z^{c}\,,}  & \medskip   &  \ensuremath{\left[Z_{a},Z_{b}\right]=\frac{1}{\ell^{2}}\epsilon_{abc}Z^{c}}\,,\\
 \ensuremath{\left[J_{a},P_{b}\right]=\epsilon_{abc}P^{c}}\,,  &  \medskip  &  \ensuremath{\left[Z_{a},P_{b}\right]=\frac{1}{\ell^{2}}\epsilon_{abc}P^{c}}\,,
\end{tabular}\ \eqlabel{adsl}
\end{equation}
where $\epsilon_{abc}$ is the Levi-Civita symbol, fullfiling $\epsilon_{012}=1$, and $a=0,1,2$ is a Lorentz index.
This algebra can be shown to be equivalent to the direct sum of three
copies of the Lorentz algebra. Indeed, three commuting sets of $\mathfrak{so}(2,1)$ generators
\begin{equation}\eqlabel{3lorentz}
\left[J_{a}^{\pm},J_{b}^{\pm}\right]= \epsilon_{abc}J^{\pm c}\,, \quad \quad
\left[\hat{J}_{a},\hat{J}_{b}\right] = \epsilon_{abc}\hat{J}^{c}\,,
\end{equation}
reproduce \eqref{adsl} considering the redefinitions
\begin{eqnarray}
Z_{a} & = & \frac{1}{\ell^{2}}\left(J_{a}^{+}+J_{a}^{-}\right)\,,\nonumber \\
P_{a} & = & \frac{1}{\ell}\left(J_{a}^{+}-J_{a}^{-}\right)\,,\\
J_{a} & = & \hat{J}_{a}+J_{a}^{+}+J_{a}^{-}\,.\nonumber
\end{eqnarray}
On the other hand, the algebra \eqref*{adsl} can be written as the direct sum $\mathfrak{so}(2,2)\oplus\mathfrak{so}(2,1)$, reason why it is usually referred to as AdS-Lorentz algebra. In fact, one could define
\begin{eqnarray}
\tilde{J}_{a} & = & \ell^{2}Z_{a}\,,\nonumber \\
\tilde{P}_{a} & = & P_{a}\,,\eqlabel{tilde}\\
\tilde{Z}_{a} & = & J_{a}-\ell^{2}Z_{a}\,,\nonumber
\end{eqnarray}
so that a direct sum of the Lorentz and AdS is made manifest
appears.

An alternative procedure to obtain the AdS-Lorentz symmetry is
through the semigroup expansion method \cite{Izaurieta:2006zz}.
As was shown in \cite{Diaz:2012zza}, the AdS-Lorentz algebra
can be seen as an $S$-expansion of the AdS algebra. As a consequence,
the non-vanishing components of an invariant tensor for the AdS-Lorentz
algebra can be obtained using Theorem
VII.2 of \cite{Izaurieta:2006zz}. Indeed, the relevant invariant tensors of rank
2 are given by
\begin{eqnarray}
\left\langle J_{a}J_{b}\right\rangle =\mu_{0}\eta_{ab}\,, & \medskip  & \left\langle P_{a}P_{b}\right\rangle =\frac{\mu_{2}}{\ell^{2}}\eta_{ab}\,,\nonumber \\
\left\langle J_{a}P_{b}\right\rangle =\frac{\mu_{1}}{\ell}\eta_{ab}\,, & \medskip  & \left\langle Z_{a}Z_{b}\right\rangle =\frac{\mu_{2}}{\ell^{4}}\eta_{ab}\,,\eqlabel{eq:invt}\\
\left\langle J_{a}Z_{b}\right\rangle =\frac{\mu_{2}}{\ell^{2}}\eta_{ab}\,, & \medskip  & \left\langle Z_{a}P_{b}\right\rangle =\frac{\mu_{1}}{\ell^{3}}\eta_{ab}\,,\nonumber
\end{eqnarray}
where $\mu_{0}$, $\mu_{1}$ and $\mu_{2}$ can be redefined as
\begin{eqnarray}
\mu_{0}\rightarrow\alpha_{0}\,, & \mu_{1}\rightarrow\alpha_{1}\ell\,, & \mu_{2}\rightarrow\alpha_{2}\ell^{2}\,,\eqlabel{redef}
\end{eqnarray}
such that the flat limit $\ell\rightarrow\infty$ is well defined
and leads to the invariant tensor of the Maxwell group. Here, $\alpha_{0},$$\,\alpha_{1}\text{ and }\alpha_{2}$
are real arbitrary constants. Note that applying the flat limit in \eqref{adsl}
reproduces the commutation relations of the Maxwell algebra. This limit has also been discussed in the context of supergravity \cite{Concha:2016zdb,Concha:2018jxx} and higher-spin theory~\cite{Caroca:2017izc}. As an ending remark, it is worth it to mention that supersymmetric extensions of the AdS-Lorentz algebra have been useful in order to restore supersymmetry invariance of supergravity theories with boundary~\cite{Ipinza:2016con,Banaudi:2018zmh,Concha:2018ywv}.

\subsection{Chern-Simons gravity}

Now we consider a three-dimensional CS gravity theory invariant under the algebra \eqref*{adsl}. The starting point is the CS action
\begin{equation}
I\left[A\right]=\frac{k}{4\pi}\int\limits _{\mathcal{M}}\langle A\mathrm{d}A+\frac{2}{3}\,A^{3}\rangle\,,\eqlabel{gcs}
\end{equation}
where $A$ is the gauge connection, $\left\langle \cdots\right\rangle $ denotes
the invariant trace, and $k$ the is Chern-Simons level, which is related to the Newton constant $G$ according to  $k=\frac{1}{4G}$. The connection one-form takes values in the the AdS-Lorentz algebra,
\begin{equation}
A=\omega^{a}J_{a}+e^{a}P_{a}+\sigma^{a}Z_{a}\,,\eqlabel{1f}
\end{equation}
where $\omega^{a}$ is the spin connection, $e^{a}$ is the
dreibein, and $\sigma^{a}$ is the gauge field associated with the
non-abelian $Z_{a}$ generator.
In terms of the gauge field components, the action takes the form
\begin{align}
I_{AdS-{\cal L}}= & \frac{k}{4\pi}\int\limits _{\mathcal{M}}\left[\alpha_{0}\left(\omega^{a}\mathrm{d}\omega_{a}+\frac{1}{3}\,\epsilon^{abc}\omega_{a}\omega_{b}\omega_{c}\right)+\alpha_{1}\left(2R_{a}e^{a}+\frac{2}{\ell^{2}}\,F^{a}e_{a}+\frac{1}{3\ell^{2}}\,\epsilon^{abc}e_{a}e_{b}e_{c}\right)\right.\nonumber \\
 & +\left.\alpha_{2}\left(T^{a}e_{a}+\frac{1}{\ell^{2}}\,\epsilon^{abc}e_{a}\sigma_{b}e_{c}+2R^{a}\sigma_{a}+\frac{2}{\ell^{2}}\,F^{a}\sigma_{a}\right)-\mathrm{d}\left(\alpha_{1}(\omega^{a}+\sigma^{a})e_{a}+\alpha_{2}\omega^{a}\sigma_{a}\right)\rule{0pt}{15pt}\right]\,.\eqlabel{cs}
\end{align}
Here, $R^{a}=d\omega^{a}+\frac{1}{2}\epsilon^{abc}\omega_{b}\omega_{c}$
corresponds to the Lorentz curvature two-form, $T^{a}=D_{\omega}e^{a}$
is the torsion two-form, and $F^{a}=D_{\omega}\sigma^{a}+\frac{1}{2\ell^{2}}\epsilon^{abc}\sigma_{b}\sigma_{c}$
is the curvature two-form associated to the gauge field $\sigma^{a}$.
One can see that the coupling constant $\alpha_{0}$ multiplies the usual gravitational CS term, while the Einstein-Hilbert term is
related to the coupling constant $\alpha_{1}$. Then, a natural choice
is to set $\alpha_{1}=1$. However, for the sake of generality we keep it arbitrary in our analysis. Of particular interest is the
presence of the extra field $\sigma^{a}$, which appears in the action through the coupling constants $\alpha_{1}$ and $\alpha_{2}$. This gauge
field affects the dynamics of the geometry. Indeed, the field equations coming from \eqref{cs} are given by
\begin{eqnarray}
\delta\omega^{a} & : &   \qquad0=\alpha_{0}R_{a}+\alpha_{1}\left(T_{a}+\frac{1}{\ell^{2}}\,\epsilon_{abc}\sigma^{b}e^{c}\right)+\alpha_{2}\left(F_{a}+\frac{1}{2}\,\epsilon_{abc}e^{b}e^{c}\right)\,,\nonumber \\ \medskip
\delta e^{a} & : &   \qquad0=\alpha_{1}\left(R_{a}+\frac{1}{\ell^{2}}\,F_{a}+\frac{1}{2\ell^{2}}\,\epsilon_{abc}e^{b}e^{c}\right)+\alpha_{2}\left(T_{a}+\frac{1}{\ell^{2}}\,\epsilon_{abc}\sigma^{b}e^{c}\right)\,,\eqlabel{eom}\\ \medskip
\delta\sigma^{a} & : &   \qquad0=\frac{\alpha_{1}}{\ell^{2}}\,\left(T_{a}+\frac{1}{\ell^{2}}\,\epsilon_{abc}\sigma^{b}e^{c}\right)+\alpha_{2}\left(R_{a}+\frac{1}{\ell^{2}}\,F_{a}+\frac{1}{2\ell^{2}}\,\epsilon_{abc}e^{b}e^{c}\right)\,.\nonumber
\end{eqnarray}
Let us note that, unlike the three-dimensional Maxwell CS theory studied in \cite{Concha:2018zeb}, three-dimensional GR without cosmological constant can neither be recovered by a limit procedure nor by eliminating some gauge field. However, it can be done by a combined procedure, namely, taking the limit $\ell\rightarrow\infty$ followed by setting $\sigma^a=0$ and $\alpha_2=0$. AdS CS gravity in three dimensions, on the other hand, can be obtained by solely considering the Euler-type CS term (proportional to $\alpha_{1}$) in the action \eqref*{cs} and setting $\sigma_{a}=0$, which leads to\footnote{This is a particular case the algebraic construction described in \cite{Concha:2016kdz,Concha:2016tms,Concha:2017nca}. The AdS-Lorentz symmetry belongs to a
family of algebras denoted as $\mathfrak{C}_m$ \cite{Concha:2016hbt}, which,
under certain conditions, leads to Pure Lovelock gravity.}
\begin{eqnarray}
R_{a}+\frac{1}{2\ell^{2}}\,\epsilon_{abc}e^{b}e^{c} & = & 0\,,\nonumber \\
T_{a} & = & 0\,. \eqlabel{Add}
\end{eqnarray}
Naturally, when the gauge field $\sigma_{a}$ is turned
on, the field equations are modified. Since $\alpha_0$ and $\alpha_2$ are still arbitrary, \eqref[s]{eom}
can be written as
\begin{eqnarray}
R_{a} & = & 0\,,\nonumber \\
T_{a}+\frac{1}{\ell^{2}}\,\epsilon_{abc}\sigma^{b}e^{c} & = & 0\,,\eqlabel{eomf}\\
F_{a}+\frac{1}{2}\,\epsilon_{abc}e^{b}e^{c} & = & 0\,.\nonumber
\end{eqnarray}

\newsection{Solutions}
\label{solutions}

In this section we study solutions of the field equations \eqref*{eomf}. To this purpose
we will deal with two different representations of the Minkowski metric.
First, we focus on the stationary solutions introduced in~\cite%
{Hoseinzadeh:2014bla}, where the Minkowski metric is taken to be on its diagonal form $\bar{\eta}=$diag$%
(-1,1,1)$. To avoid confusion, gauge fields whose indices are raised and lowered with this metric, will be denoted with bars. Subsequently, we study solution in the BMS gauge, where the Minkowski metric is written in the light-cone representation
	\begin{equation}
	\eta_{ab}=\left(\begin{array}{ccc}
	0 & 1 & 0\\
	1 & 0 & 0\\
	0 & 0 & 1
	\end{array}\right)\,. \eqlabel{eta}
	\end{equation}

\subsection{Stationary solutions and gauge fixing}
\label{stationary}

Now we consider the BTZ-type solution of the field equations \eqref*{eomf}. We
also calculate the conserved charges of the theory, which, as we will see,
are modified with respect to the pure gravity case by the presence of the gauge field $\sigma ^{a}$. This solution
was first presented in~\cite{Hoseinzadeh:2014bla}.  The ADM form of the
metric is
\begin{equation}
ds^{2}=-N^{2}dt^{2}+\frac{dr^{2}}{N^{2}}+r^{2}\left( d\varphi +N_{\varphi
}dt\right) ^{2}\,,  \eqlabel{ADM}
\end{equation}%
where
\begin{equation}
N^{2}=-M+\frac{J^{2}}{4r^{2}}+\frac{r^{2}}{\ell ^{2}}\,,\qquad N_{\varphi }=-%
\frac{J}{2r^{2}}\,.
\eqlabel{N_and_Nphi}
\end{equation}%
Here $M$ and $J$ are integration constants. The dreibein one-forms can be chosen as
\begin{align}
\bar{e}^{0}& =Ndt\,,  \notag \\
\bar{e}^{1}& =N^{-1}dr\,,  \eqlabel{btz1} \\
\bar{e}^{2}& =r\left( d\varphi +N_{\varphi }dt\right) \,.  \notag
\end{align}%
With this choice, the solution found in~\cite{Hoseinzadeh:2014bla} for the
spin-connection one-form and the non-abelian gauge field $\bar{\sigma}$
reads
\begin{align}
\bar{\omega}^{0}& =aB dt+C dr+B d\varphi \,,  \notag  \medskip  \medskip \\
\bar{\omega}^{1}& =aHdt+E dr+Hd\varphi \,,  \eqlabel{btz2} \medskip \\
\bar{\omega}^{2}& =aKdt+Fdr+Kd\varphi \,,  \notag
\end{align}%
\begin{align}
\bar{\sigma}^{0}& =-\ell ^{2}\left[ aB dt+C dr+(B -N)d\varphi \right]
\,,  \notag  \medskip \\
\bar{\sigma}^{1}& =-\ell ^{2}\left[ aHdt+\left( E +\frac{N_{\varphi }}{N}%
\right) dr+Hd\varphi \right] \,\,,  \eqlabel{btz3} \\
\bar{\sigma}^{2}& =\,-\ell ^{2}\left[ \left( -\frac{r}{\ell ^{2}}+aK\right)
dt+Fdr+(K-r N_{\varphi })d\varphi )\right] \,,  \notag
\end{align}%
where we have defined
\begin{eqnarray}
B &=&\sqrt{K^{2}+H^{2}+b\,}\,, \medskip \notag \\
C &=&\frac{H^{\prime }+B F}{K}\,,\medskip \eqlabel{BCE} \\
E &=&\frac{KK^{\prime }+HH^{\prime }}{KB }+\frac{HF}{K}\,.  \notag
\end{eqnarray}%
Note that the solution depends on three functions $F(r)$, $K(r)$
and $H(r)$ and two additional constants $a$ and $b$.
We choose to fix these functions as%
\begin{equation}
F=\frac{r N_{\varphi }}{N^{2}}+\frac{%
N^{\prime }}{N}\,,\quad H=\frac{b+G^{2}-N^{2}}{2N}\,,\quad K=r N_{\varphi }
\eqlabel{gauge_fix}
\end{equation}%
since, as we will see in Subsection \ref{bmsgauge} this choice allows one to connect the results presented here with the ones written in a BMS-like ansatz for the theory.

It is important to remark that the presence of the arbitrary functions in \eqref{BCE} is due to a residual
gauge symmetry which has not been fixed yet. Indeed, the BTZ ansatz for the
metric (3.4) is obtained by assuming that it is stationary, axially symmetric
so that it has $\partial_{t}$ and $\partial_{\phi}$ as the Killing vectors.
For the field $\bar{\omega}$ and $\bar{\sigma}^{a}$ the same symmetries of the
metric are assumed, reason why \eqref[s]{btz2}--\eqref*{BCE} have radial dependence only.
This already fixes a considerable part of the gauge symmetry. However, there
is still a residual symmetry that can be used to fix the arbitrary functions $F$, $H$ and $K$ (and therefore they are pure gauge). In other words, one can look
for restricted gauge transformations $\delta_{\Lambda} A=d\Lambda+\left[  A,\Lambda\right]$, with $\Lambda$  a gauge parameter taking values in the AdS-Lorentz algebra, that map a given solution with
Killing vectors $\partial_{t}$ and $\partial_{\phi}$ in another solution
having the same Killing vectors. By doing this, one can completely identify
the gauge transformation $A^{\prime}\rightarrow A=A^{\prime}+\delta_{\Lambda}A^{\prime}$ such that, starting from the solution (3.6) with an arbitrary choice of the functions  $F$, $H$ and $K$, one ends up with a solution of the same form satisfying (3.8).

Let us now focus on the calculation of the Noether charges. As it is very
well-known, in CS gravity these are given by~\cite{Julia:1998ys, Feng:1999mk}%
\begin{equation}
Q[\xi ]=\frac{k }{4\pi }\,\int\limits_{\partial \Sigma }\langle A\iota
_{\xi }A\rangle \,,
\end{equation}%
where $\xi ^{i}$ are the asymptotic Killing vectors, and the charges are
evaluated at the asymptotic infinity and at a constant time slice $\partial
\Sigma $. Considering bared gauge fields in \eqref{1f}, it becomes
\begin{eqnarray}
Q[\xi ] &=&\frac{k }{4\pi }\lim_{r\rightarrow \infty
}\int\limits_{0}^{2\pi }d\varphi \,\xi ^{i}\left[ \alpha _{0}\bar{\omega}%
_{\varphi }^{a}\bar{\omega}_{ai}+ \alpha _{1}\left(\bar{\omega}_{\varphi }^{a}\bar{e}_{ai}+%
\bar{e}_{\varphi }^{a}\bar{\omega}_{ai}+\frac{1}{\ell ^{2}}(\bar{e}_{\varphi
}^{a}\bar{\sigma} _{ai}+\bar{\sigma} _{\varphi }^{a}\bar{e}_{ai})\right) \right.  \notag \\
&&\left. +\alpha _{2}\left( \bar{\omega}_{\varphi }^{a}\bar{\sigma}_{ai}+%
\bar{e}_{\varphi }^{a}\bar{e}_{ai}+\bar{\sigma}_{\varphi }^{a}\bar{\omega}%
_{ai}+\frac{1}{\ell ^{2}}\bar{\sigma} _{\varphi }^{a}\bar{\sigma} _{ai}\right) \right]
\,.  \eqlabel{CS charge}
\end{eqnarray}%
Using this formula, we compute the
conserved charges of the solution \eqref*{btz1}--\eqref*{btz3}, associated with
asymptotic invariance under time translations $\xi =\partial _{t}$ (mass $m)$
and rotations $\xi =\partial _{\varphi }$ (angular momentum $j$). These read
\begin{eqnarray}
m &\equiv &Q[\partial _{t}]=\frac{k}{2}[-\alpha _{0}ab+ \alpha _{1}M+\alpha _{2}(\ell
^{2}ab-J)]\,,  \notag \\
j &\equiv &Q[\partial _{\varphi }]=\frac{k}{2}[-\alpha _{0}b- \alpha _{1}J+\alpha
_{2}\ell ^{2}(b+M)]\,.  \eqlabel{mj}
\end{eqnarray}%
As in the Maxwell case~\cite{Concha:2018zeb}, the gauge field $\sigma^a$ contributes to the
mass and angular momentum of the solution, and therefore, modifies the
asymptotic sector.

\subsection{Solutions in the BMS gauge}
\label{bmsgauge}

As was previously discussed, there is a basis where the AdS-Lorentz symmetry
can be written as the direct sum $\mathfrak{so}(2,2)\oplus \mathfrak{so}(2,1)$. Thus, a trick can be used to define suitable boundary conditions for the theory: we can go to the direct product basis \eqref*{tilde}, where the torsionless fields $\tilde{e}^{a}$ and $\tilde{\omega}^{a}$ can be set to obey standard pure gravity boundary conditions and $\tilde{\sigma}^{a}$ can be considered simply as a flat Lorentz connection. Subsequently we can go back to the original AdS-Lorentz basis. This is the strategy that we will adopt in the following analysis.

For later convenience, the asymptotically AdS geometries in the direct product basis \eqref*{adsl} will be described in three-dimensional BMS gauge~\cite{Barnich:2012aw,Barnich:2013yka}, where the manifold is parameterized by the local
coordinates $x^{\mu }=(u,r,\phi )$. The metric takes the form
\begin{equation}
ds^{2}=\left( \mathcal{M}(u,\phi)-\frac{r^{2}}{\ell ^{2}}\right) du^{2}-2dudr+%
\mathcal{N}(u,\phi)d\phi du+r^{2}d\phi ^{2}\,,  \eqlabel{BMS1}
\end{equation}
were $u$ is the retarded time coordinate and the boundary is located at $%
r=const$.
Considering the off-diagonal Minkowski metric \eqref*{eta}, this can be written in
terms of the dreibein as
\begin{equation}
ds^{2}=2\tilde{e}^{0}\tilde{e}^{1}+\left( \tilde{e}^{2}\right) ^{2}\,.
\eqlabel{BMS3}
\end{equation}%
Therefore, the dreibein one-forms can be chose as
\begin{equation}
\begin{tabular}{ll}
$\tilde{e}^{0}$ & $=-dr+\dfrac{1}{2}\mathcal{M}(u,\phi) du+\dfrac{%
1}{2}\mathcal{N}(u,\phi) d\phi -\dfrac{r^{2}}{2\ell ^{2}}du\,,$
\\
$\tilde{e}^{1}$ & $=du\,,$ \\
$\tilde{e}^{2}$ & $=rd\phi \,,$%
\end{tabular}%
\end{equation}%
In the basis, where the AdS-Lorentz is manifestly written as $\mathfrak{so}(2,2)\oplus \mathfrak{so}(2,1)$, the spin connection $\tilde{\omega}^a$ is torsionless. Therefore it has the form
\begin{equation}
\begin{tabular}{ll}
$\tilde{\omega}^{0}$ & $=\dfrac{1}{2}\mathcal{M}(u,\phi) d\phi +%
\dfrac{1}{2\ell ^{2}}\mathcal{N}(u,\phi) du-\dfrac{r^{2}}{2\ell
^{2}}d\phi \,,$ \\
$\tilde{\omega}^{1}$ & $=d\phi \,,$ \\
$\tilde{\omega}^{2}$ & $=\dfrac{r}{\ell ^{2}}du\,,$ \\
\end{tabular}%
\end{equation}%
Finally, for the gauge firld $\tilde{\sigma}^{a}$ we consider
\begin{align}
\tilde{\sigma}^{0}& =\dfrac{1}{2}\left(\mathcal{M}(u,\phi)-\dfrac{1}{\ell^2}\mathcal{R}(u,\phi)\right)d\phi \,,  \notag \\
\tilde{\sigma}^{1}& =d\phi \,,  \eqlabel{BMS6} \\
\tilde{\sigma}^{2}& =0\,.  \notag
\end{align}
As explained above, in the basis \eqref*{tilde}, the field $\tilde\sigma^a$ gets decoupled and the solutions for the fields  $\tilde{e}^{a}$ and $\tilde{\omega}^{a}$ are the ones of three-dimensional GR with negative cosmological constant . Thus, we can use the well-known
results for the metric fields in asymptotically AdS three-dimensional Einstein gravity. In the BMS gauge this means (see for instance \cite{Barnich:2013yka}),
\begin{equation}
\mathcal{\dot{M}}(u,\phi )=\frac{1}{\ell ^{2}}\mathcal{N}^{\prime }(u,\phi )\;,%
\quad\quad\mathcal{\dot{N}}(u,\phi )=\mathcal{M}^{\prime
}(u,\phi )\;,
\eqlabel{BMS7}
\end{equation}%
where prime and dot denote derivative with respect to
the coordinate $\phi $ and $u$, respectively, and whose solution is
\begin{equation}\eqlabel{Lpm}
\mathcal{M}=\mathcal{L}^+ + \mathcal{L}^-\;,\quad \mathcal{N}=\ell\left(\mathcal{L}^+ - \mathcal{L}^-\right)\;,
\end{equation}
with $\mathcal{L}^\pm = \mathcal{L}^\pm(x^\pm)$ and $x^\pm = \phi \pm \frac{1}{\ell} u$.

The field equation for the field $\tilde\sigma^a$, on the other hand, leads to the extra condition
\begin{equation}\eqlabel{eqr}
\dot{\mathcal{R}}(u,\phi )=\mathcal{N}^{\prime }(u,\phi )\,,
\end{equation}
which yields
\begin{equation}
\mathcal{R}=\ell^2\left(\mathcal{L}^+ + \mathcal{L}^--2\mathcal{L}\right)\,,\quad\quad\mathcal{L}=\mathcal{L}(\phi).  \eqlabel{BMS8}
\end{equation}

Since we are interested in the basis where the Maxwell symmetry is found as a
flat limit of the AdS-Lorentz one, we need a BMS-like solution  for
our original gauge fields $(e^{a},\omega ^{a},\sigma ^{a})$. \ It is
 straightforward to show by means of \eqref{tilde} that these fields are related to the previous ones $(\tilde{%
e}^{a},\tilde{\omega}^{a},\tilde{\sigma}^{a})$ as follows:
\begin{equation}
e^{a}=\tilde{e}^{a},\text{ \ \ \ \ }\omega ^{a}=\tilde{\sigma}^{a},\text{ \
\ \ }\sigma ^{a}=\ell ^{2}(\tilde{\omega}^{a}-\tilde{\sigma}^{a})\,.
\end{equation}%
Therefore, we have that
the field equations \eqref*{eomf} coming from the AdS-Lorentz CS gravity action are
satisfied by the following components of the gauge fields,

\begin{equation} \eqlabel{bmsg}
\begin{array}{llllll}
e^{0} =-dr+\dfrac{1}{2}\left(\mathcal{M}-\dfrac{r^{2}}{\ell ^{2}}\right) du+\dfrac{1}{2}\mathcal{N} d\phi\,,\quad & e^{1} =du\,, \quad & e^{2}  =rd\phi \,,\\
\omega ^{0}  =\dfrac{1}{2}\left(\mathcal{M}-\dfrac{\mathcal{R}}{\ell^2}\right) d\phi \,,\quad & \omega ^{1}  =d\phi \,, \quad & \omega ^{2}  =0\,,\\
\sigma ^{0}  =\dfrac{1}{2}\mathcal{N} du+\dfrac{1}{2}\left( \mathcal{R}-r^{2}\right) d\phi \,,\quad & \sigma ^{1}  =0\,, \quad & \sigma ^{2}  =rdu\,.
\end{array}
\end{equation}

These are the boundary conditions for the Maxwell connection \eqref*{1f} that we will adopt for our analysis of the asymptotic symmetry algebra.
Note that in the limit $\ell \rightarrow \infty $, the solution \eqref*{bmsg} reduces to the Maxwell case presented in~\cite{Concha:2018zeb}. Also, when taking this limit in \eqref[s]{BMS7} and \eqref{eqr}, they lead to $\mathcal{M=M(\phi
)}$, $\mathcal{N}=u\mathcal{M}^{\prime }+\mathcal{J}(\phi )$ and $\mathcal{R} =%
\frac{u^{2}}{2}\mathcal{M}^{\prime \prime }+u\mathcal{J}^{\prime }+\mathcal{Z%
}(\phi )$, which is precisely the result obtained in \cite{Concha:2018zeb}.

The boundary conditions \eqref*{bmsg} contain the stationary solution studied in Subsection \ref{stationary}, when the functions $\mathcal{M}$, $\mathcal{N}$ and
$\mathcal{R} $ are constant, namely,
\begin{equation}
\mathcal{M}\left( u,\phi \right) =M\,,\text{ \ \ \ \ }\mathcal{N}\left(
u,\phi \right) =-J\,,\text{ \ \ \ }\mathcal{R} (u,\phi )=\ell ^{2}(b+M)\,.
\eqlabel{BMS13}
\end{equation}
The first and second expressions in \eqref{BMS13} are straightforward, while the third one can be obtained by proceeding in a similar way as done in \cite{Concha:2018zeb}. The trick consists in using the change of variables familiar from the pure gravity case \cite{Barnich:2012xq}, $t=u+f(r)\,,\; \varphi = \phi+g(r)$ with $f^{\prime}= N^{-2}$ and $g^{\prime}=-N^\varphi f'$, and define the the following matrix
\begin{equation}
K_{\ b}^{a}=e_{\mu }^{a}\bar{e}_{b}^{\mu }=\left(
\begin{array}{ccc}
-\frac{1}{2N}\,\left( N_{\varphi }^{2}r^{2}+N^{2}\right) & \frac{1}{2N}%
\,\left( N_{\varphi }^{2}r^{2}-N^{2}\right) & rN_{\varphi } \\
N^{-1} & -N^{-1} & 0 \\
-rN_{\varphi }N^{-1} & rN_{\varphi }N^{-1} & 1%
\end{array}%
\right) \,,  \eqlabel{KK}
\end{equation}
where $N$ and $N_{\varphi }$ given by \eqref{N_and_Nphi }. This matrix can be used to map the gauge field $\bar\sigma^a$ in the diagonal basis $\bar{e}^a$ to the corresponding one in the off-diagonal basis $e^a$ as $\sigma _{\mu }^{a}=K_{\ b}^{a}\bar{\sigma}_{\mu }^{b}$. Then, after a simple calculation one can see that the stationary solution given in \eqref[s]{ADM}--\eqref*{gauge_fix} is recovered from \eqref*{bmsg} when \eqref{KK} is implemented

\newsection{Asymptotic Symmetry}
\label{asym}

In order to compute the asymptotic symmetry algebra associated to the AdS-Lorentz CS
gravity, we have to consider suitable fall-off conditions for the gauge
fields at infinity. Using \eqref{bmsg}, it is direct to evaluate the gauge connection $A$ in the BMS gauge:
\begin{eqnarray}
A &=&\left( \frac{1}{2}\mathcal{N}\left( u,\phi \right) du+\frac{1}{2}\mathcal{R}
\left( u,\phi \right) d\phi -\frac{r^{2}}{2}d\phi \right) Z_{0}+rduZ_{2}
\nonumber \\
&&+\left( -dr+\frac{1}{2}\mathcal{M}\left( u,\phi \right) du+\frac{1}{2}%
\mathcal{N}\left( u,\phi \right) d\phi -\frac{r^{2}}{2\ell ^{2}}du\right)
P_{0}+duP_{1}+rd\phi P_{2}  \eqlabel{connection} \\
&&+\left( \frac{1}{2}\mathcal{M}\left( u,\phi \right) -\frac{1}{2\ell ^{2}}%
\mathcal{R} \left( u,\phi \right) \right) d\text{$\phi $}J_{0}+d\phi J_{1}\,.
\nonumber
\end{eqnarray}
Furthermore, the radial dependence of the connection \eqref*{connection}, can be gauged away by an
appropriate gauge transformation%
\begin{equation}
A=h^{-1}dh+h^{-1}ah\,,
\end{equation}%
where $h=e^{-rP_{0}}$. Then, using the Baker-Campbell-Hausdorff formula and
the identity $h^{-1}dh=-dr\,P_{0}$, we obtain the following asymptotic field:
\begin{eqnarray}
a &=&\left( \frac{1}{2}\mathcal{N}\left( u,\phi \right) du+\frac{1}{2}\mathcal{R}
\left( u,\phi \right) d\phi \right) Z_{0}+\left( \frac{1}{2}\mathcal{M}%
\left( u,\phi \right) du+\frac{1}{2}\mathcal{N}\left( u,\phi \right) d\phi
\right) P_{0}  \notag \\
&&+duP_{1}+\left( \frac{1}{2}\mathcal{M}\left( u,\phi \right) -\frac{1}{%
2\ell ^{2}}\mathcal{R} \left( u,\phi \right) \right) d\text{$\phi $}J_{0}+d\phi
J_{1}\,.  \eqlabel{aadslor}
\end{eqnarray}

\subsection{Residual gauge transformations}
The next step consists in finding the residual gauge transformations, $\delta_\Lambda A=d\Lambda +%
\left[ A,\Lambda \right] $, that preserve our boundary conditions \eqref*%
{connection}. To this purpose, let us consider gauge parameters of the form
\begin{equation}
\Lambda =h^{-1}\lambda h\,,\qquad \lambda =\chi ^{a}\left( u,\phi \right)
J_{a}+\varepsilon ^{a}\left( u,\phi \right) P_{a}+\gamma ^{a}\left( u,\phi
\right) Z_{a}\,.  \eqlabel{lambda2}
\end{equation}%
Then, gauge transformations of the connection $A$ with gauge
parameter $\Lambda $ imply $r$-independent gauge transformations of $a$ with
gauge parameter $\lambda $. Therefore, the variation of the asymptotic field \eqref*
{aadslor} reads
\begin{eqnarray}
\delta _{\lambda }a &=&\frac{1}{2}\,\left( \delta _{\lambda }\mathcal{N}%
\left( u,\phi \right) du+\delta _{\lambda }\mathcal{R} \left( u,\phi \right)
d\phi \right) \,Z_{0}+\frac{1}{2}\left( \text{$\delta _{\lambda }$}\mathcal{M%
}\left( u,\phi \right) du+\text{$\delta _{\lambda }$}\mathcal{N}\left(
u,\phi \right) d\phi \right) \,P_{0}  \notag \\
&&+\frac{1}{2}\left( \text{$\delta _{\lambda }$}\mathcal{M}\left( u,\phi
\right) d\text{$\phi -\frac{1}{\ell ^{2}}\delta _{\lambda }\mathcal{R} \left(
u,\phi \right) d\phi $}\right) \,J_{0}\,,  \eqlabel{aphi2}
\end{eqnarray}%
and has to be equal to a gauge transformation of the form
\begin{equation}
\delta _{\lambda }a=d\lambda +\left[ a,\lambda \right] \,.
\eqlabel{deltaa2}
\end{equation}%

Let us focus first on the angular component of the connection $a$, and replace \eqref[s]{aadslor} and \eqref*{lambda2 } in \eqref{deltaa2 }. We find
\begin{eqnarray}
\delta_{\lambda} a_{\phi } &=&\left( \gamma ^{0\prime }-\frac{\text{$\mathcal{R} $}}{2}%
\,\chi ^{2}-\frac{\mathcal{N}}{2}\,\varepsilon ^{2}-\frac{\mathcal{M}}{2}%
\,\gamma ^{2}\right) \,Z_{0}+\left( \gamma ^{1\prime }+\gamma ^{2}\right)
\,Z_{1}\nonumber\\&&+\left( \gamma ^{2\prime }+\frac{\text{$\mathcal{R} $}}{2}\,\chi ^{1}+%
\frac{\mathcal{N}}{2}\,\varepsilon ^{1}+\frac{\mathcal{M}}{2}\,\gamma
^{1}-\gamma ^{0}\right) \,Z_{2}  \nonumber \\
&&+\left[ \varepsilon ^{0\prime }-\frac{\mathcal{N}}{2}\,\left( \frac{\gamma
^{2}}{\ell ^{2}}+\chi ^{2}\right) -\frac{\mathcal{M}}{2}\,\varepsilon
^{2}\right] \,P_{0}+\left( \varepsilon ^{1\prime }+\varepsilon ^{2}\right)
\,P_{1} \eqlabel{var1}\\
&&+\left[ \varepsilon ^{2\prime }+\frac{\mathcal{N}}{2}\,\left( \frac{\gamma
^{1}}{\ell ^{2}}+\chi ^{1}\right) +\frac{\mathcal{M}}{2}\,\varepsilon
^{1}-\varepsilon ^{0}\right] \,P_{2}+\left[ \chi ^{0\prime }+\frac{1}{2}%
\,\left( \frac{\mathcal{R} }{\ell ^{2}}-\mathcal{M}\right) \,\chi ^{2}\right]
\,J_{0}  \nonumber \\
&&+\left( \chi ^{1\prime }+\chi ^{2}\right) \,J_{1}+\left[ \text{$\chi $}%
^{2\prime }+\frac{1}{2}\,\left( \mathcal{M-}\frac{\mathcal{R} }{\ell ^{2}}\right)
\,\chi ^{1}-\chi ^{0}\right] \,J_{2}\,.  \nonumber
\end{eqnarray}%
From \eqref[s]{aphi2} and \eqref*{var1}, one immediately sees that the arbitrary functions appearing in \eqref*{connection} must satisfy the following relations:
\begin{eqnarray}
\delta _{\lambda }\mathcal{M}+\frac{\delta _{\lambda }\mathcal{R} }{\ell ^{2}}
&=&2\chi ^{0\prime }-\mathcal{M}\chi ^{2}+\frac{\mathcal{R} }{\ell ^{2}}\chi
^{2}\,,  \notag \\
\delta _{\lambda }\mathcal{N} &=&2\varepsilon ^{0\prime }-\mathcal{N\,}%
\left( \frac{\gamma ^{2}}{\ell ^{2}}+\chi ^{2}\right) -\mathcal{M}%
\varepsilon ^{2}\,, \eqlabel{deltafunctions1}\\
\delta _{\lambda }\mathcal{R} &=&2\gamma ^{0\prime }-\mathcal{R} \chi ^{2}-\mathcal{N}%
\varepsilon ^{2}-\mathcal{M}\gamma ^{2}\,,\notag
\end{eqnarray}%
while the gauge parameters satisfy%
\begin{equation}\eqlabel{egp1}
\begin{tabular}{ll}
$\gamma ^{1\prime }+\gamma ^{2}=0\,,$ & $\gamma ^{2\prime }+\frac{\text{$%
\mathcal{R} $}}{2}\,\chi ^{1}+\frac{\mathcal{N}}{2}\,\varepsilon ^{1}+\frac{%
\mathcal{M}}{2}\,\gamma ^{1}-\gamma ^{0}=0\,,$ \medskip \\
$\varepsilon ^{1\prime }+\varepsilon ^{2}=0\,,$ & $\varepsilon ^{1\prime }+%
\frac{\mathcal{N}}{2}\left( \frac{\gamma ^{1}}{\ell ^{2}}+\chi ^{1}\right) +%
\frac{\mathcal{M}}{2}\varepsilon ^{1}-\varepsilon ^{0}=0\,,$ \medskip \\
$\chi ^{1\prime }+\chi ^{2}=0\,,$ & $\text{$\chi $}^{2\prime }+\frac{%
\mathcal{M}}{2}\,\chi ^{1}-\frac{\mathcal{R} }{2\ell ^{2}}\,\chi ^{1}-\chi
^{0}=0\,.$%
\end{tabular}%
\end{equation}%
On the other hand, the variation of the $u$-component of the gauge field $a$
is given by
\begin{eqnarray}
\delta _{\lambda }a_{u} &=&\left[ \dot{\gamma}^{0}-\frac{\mathcal{N}}{2}%
\,\left( \frac{\gamma ^{2}}{\ell ^{2}}+\chi ^{2}\right) -\frac{\mathcal{M}}{2%
}\,\varepsilon ^{2}\right] \,Z_{0}+\left( \dot{\gamma}^{1}+\varepsilon
^{2}\right) \,Z_{1}  \notag \\
&&+\left[ \dot{\gamma}^{2}+\frac{\mathcal{N}}{2}\,\left( \frac{\gamma ^{1}}{%
\ell ^{2}}+\chi ^{1}\right) +\frac{\mathcal{M}}{2}\,\varepsilon
^{1}-\varepsilon ^{0}\right] \,Z_{2}  \eqlabel{var2} \\
&&+\left[ \dot{\varepsilon}^{0}-\frac{\mathcal{M}}{2}\,\left( \frac{\gamma
^{2}}{\ell ^{2}}+\chi ^{2}\right) -\frac{\mathcal{N}}{2\ell ^{2}}\varepsilon
^{2}\right] \,P_{0}+\left( \dot{\varepsilon}^{1}+\chi ^{2}+\frac{\gamma ^{2}%
}{\ell ^{2}}\right) \,P_{1}  \notag \\
&&+\left[ \dot{\varepsilon}^{2}+\frac{\mathcal{M}}{2}\,\left( \frac{\gamma
^{1}}{\ell ^{2}}+\chi ^{1}\right) +\frac{\mathcal{N}}{2\ell ^{2}}%
\,\varepsilon ^{1}-\chi ^{0}-\frac{\gamma ^{0}}{\ell ^{2}}\right] \,P_{2}+%
\dot{\chi}^{a}J_{a}\,.  \nonumber
\end{eqnarray}%
Comparing \eqref{var2 }
with the $u$-component of \eqref{aphi2 }, we obtain the conditions
\begin{equation} \eqlabel{deltafunctions2}
\begin{array}{lcl}
\delta _{\lambda }\mathcal{M} &=&2\dot{\varepsilon}^{0}-\mathcal{M}\,\left(
\frac{\gamma ^{2}}{\ell ^{2}}+\chi ^{2}\right) -\frac{\mathcal{N}}{\ell ^{2}}%
\varepsilon ^{2}\,,\\[5pt]
\delta _{\lambda }\mathcal{N} &=&2\dot{\gamma}^{0}-\mathcal{N}\,\left( \frac{%
\gamma ^{2}}{\ell ^{2}}+\chi ^{2}\right) -\mathcal{M}\,\varepsilon ^{2}\,,
\end{array}%
\end{equation}
together with $\dot{\chi}^{a}=0$. Furthermore, \eqref{var2 } implies that the components of the gauge parameter $\lambda$ satisfy
\begin{equation}\eqlabel{egp2}
\begin{tabular}{ll}
$\dot{\gamma}^{1}+\varepsilon ^{2}=0\,,$ & $\dot{\gamma}^{2}+\frac{\mathcal{N%
}}{2}\,\left( \frac{\gamma ^{1}}{\ell ^{2}}+\chi ^{1}\right) +\frac{\mathcal{%
M}}{2}\,\varepsilon ^{1}-\varepsilon ^{0}=0\,,$ \medskip \\
$\dot{\varepsilon}^{1}+\chi ^{2}+\frac{\gamma ^{2}}{\ell ^{2}}=0\,,$ & $\dot{%
\varepsilon}^{2}+\frac{\mathcal{M}}{2}\,\left( \frac{\gamma ^{1}}{\ell ^{2}}%
+\chi ^{1}\right) +\frac{\mathcal{N}}{2\ell ^{2}}\,\varepsilon ^{1}-\chi
^{0}-\frac{\gamma ^{0}}{\ell ^{2}}=0\,.$%
\end{tabular}%
\end{equation}
Equations \eqref*{egp1} and \eqref*{egp2} can be solved in terms of $\chi ^{1}=Y\left( \phi \right) $, $\varepsilon
^{1}=f\left(u,\phi \right) $ and $\gamma ^{1}=h\left(u, \phi \right) $, where $Y$, $f$ and $h$ are arbitrary functions of their arguments. After some calculations one gets
\begin{equation}
\begin{array}{llllll}
\chi^{0} & =\left(\frac{\mathcal{M}}{2}-\frac{\mathcal{R}}{2 \ell^2}\right)\,Y-Y^{\prime\prime}\,,\medskip & \varepsilon^{0} & =\left(\frac{\mathcal{M}f}{2}+\frac{\mathcal{N}Y}{2}+\frac{\mathcal{N}h}{2\ell^2}\right)-f^{\prime\prime}\,, & \gamma^{0} & =\left(\frac{\mathcal{M}h}{2}+\frac{\mathcal{N}f}{2}+\frac{\mathcal{R} Y}{2}\right)-h^{\prime\prime},\\
\chi^{2} & =-Y^{\prime}\,,\medskip & \varepsilon^{2} & =-f^{\prime}\,, & \gamma^{2} & =-h^{\prime}\,,\\
\chi^{1} & =Y\,, & \varepsilon^{1} & =f\,, & \gamma^{1} & =h\,.
\end{array}
\end{equation}
This leads to the following solution for \eqref*{deltafunctions1} and \eqref*{deltafunctions2} for the transformation laws of $\mathcal{M}$, $\mathcal{N}$ and $%
\mathcal{R}$:
\begin{eqnarray}
\delta \mathcal{M} &=&\mathcal{M}^{\prime }Y+2\mathcal{M}Y^{\prime
}-2Y^{\prime \prime \prime }+\frac{2}{\ell ^{2}}\,\left( \mathcal{M}h
^{\prime }+\mathcal{N}f ^{\prime }-h ^{\prime \prime
\prime }+\frac{h\mathcal{M}^{\prime }}{2}+\frac{f%
\mathcal{N}^{\prime }}{2}\right) \,,  \notag  \eqlabel{deltam} \\
\delta \mathcal{N} &=&\mathcal{M}^{\prime }f+2\mathcal{M}%
f ^{\prime }-2f ^{\prime \prime \prime }+\mathcal{N}%
^{\prime }Y+2\mathcal{N}Y^{\prime }+\frac{1}{\ell ^{2}}\left( 2\mathcal{N}%
h ^{\prime }+h \mathcal{N}^{\prime }\right) \,,
\eqlabel{deltan}  \\ \medskip
\delta \mathcal{R} &=&\mathcal{M}^{\prime }h+2\mathcal{M}h
^{\prime }-2h ^{\prime \prime \prime }+\mathcal{N}^{\prime
}f+2\mathcal{N}f ^{\prime }+\mathcal{R} ^{\prime
}Y+2\mathcal{R} Y^{\prime }\,.  \notag  \eqlabel{deltalpha}
\end{eqnarray}

The asymptotic structure of AdS-Lorentz CS gravity with boundary conditions \eqref*{connection} is contained in the
transformation laws of the functions $\mathcal{M},\mathcal{N}$ and $\mathcal{R} $%
. Indeed, the charge algebra of the AdS-Lorentz theory can be computed
following the Regge-Teitelboim approach~\cite{Regge:1974zd}. In particular, the variation of
the charge generators in a three-dimensional CS theory is given by \cite{Banados:1998gg}:
\begin{equation}\eqlabel{Qcs}
\delta Q\left[ \lambda \right] =\frac{k}{2\pi }\int d\phi \left\langle
\lambda \delta a_{\phi }\right\rangle \,,
\end{equation}%
where the non-vanishing components of the invariant tensor for the
AdS-Lorentz algebra are given by \eqref{eq:invt}. Then, using \eqref*{aphi2}, we find
\begin{align}
\delta Q\left[ Y,f,h\right] =&-\frac{k}{4\pi }\int d\phi \,\left[ Y\,\left(
\alpha _{0}\delta \mathcal{M}+\left( \alpha _{2}-\frac{\alpha _{0}}{\ell ^{2}%
}\right) \delta \mathcal{R} +\alpha _{1}\delta \mathcal{N}\right)\right. \nonumber\\
&+\left. f\,\left(
\alpha _{2}\delta \mathcal{N}+\alpha _{1}\delta \mathcal{M}\right)+ h\,\left( \alpha _{2}\delta \mathcal{M}+\frac{\alpha _{1}\delta \mathcal{N}%
}{\ell ^{2}}\right) \right] \,.
\end{align}
Since this expression is linear in the
variation of the functions $\mathcal{M}$, $\mathcal{N}$ and $\mathcal{R}$, it can be directly integrated to give
\begin{align}
Q\left[ Y,f,h\right] =&-\frac{k}{4\pi }\int d\phi \,\left[ Y\,\left( \alpha
_{0}\mathcal{M}+\left( \alpha _{2}-\frac{\alpha _{0}}{\ell ^{2}}\right)
\,\mathcal{R} +\alpha _{1}\mathcal{N}\right)\right. \nonumber \\
&+\left. f\,\left( \alpha _{2}\mathcal{N}%
+\alpha _{1}\mathcal{M}\right) +h\,\left( \alpha _{2}\mathcal{M}+\frac{%
\alpha _{1}\mathcal{N}}{\ell ^{2}}\right) \right] \,.  \eqlabel{chargeQ}
\end{align}

\subsection{Charge algebra}
Following \cite{Regge:1974zd}, the Poisson algebra of the conserved charges can be evaluated by looking at ther variations under gauge transformations,
\begin{equation}\eqlabel{rt}
\delta _{\Lambda _{2}}Q\left[ \Lambda _{1}\right] =\left\{ Q\left[ \Lambda
_{1}\right] ,Q\left[ \Lambda _{2}\right] \right\} \,.
\end{equation}%
From the expression \eqref*{chargeQ }, we see that it is possible to define independent charges for each parameter $Y$, $T$ and $R$ as
\begin{eqnarray}
j[Y] &=&\frac{k}{4\pi }\int d\phi \,Y\,\left[ \alpha _{0}\mathcal{M}+\left(
\alpha _{2}-\frac{\alpha _{0}}{\ell ^{2}}\right) \,\mathcal{R} +\alpha
_{1}\mathcal{N}\right] \,, \nonumber \\
p[f] &=&\frac{k}{4\pi }\int d\phi \,f\,\left( \alpha _{2}\mathcal{N}+\alpha
_{1}\mathcal{M}\right) \,, \\
z[h] &=&\frac{k}{4\pi }\int d\phi \,h\,\left( \alpha _{2}\mathcal{M}+\frac{%
\alpha _{1}\mathcal{N}}{\ell ^{2}}\right) \,. \nonumber
\end{eqnarray}%
The Poisson brackets of these charges can be evaluated using \eqref[s]{rt} and \eqref*{deltan}, leading to
\begin{eqnarray}
\left\{ j[Y_{1}],j[Y_{2}]\right\} &=&j\left[ [Y_{1},Y_{2}]\right] -\frac{%
k\alpha _{0}}{2\pi }\int d\phi Y_{1}Y_{2}^{\prime \prime \prime }\,,  \notag \\
\left\{ j[Y],p[f]\right\} &=&p\left[ [Y,f]\right] -\frac{k\alpha_1 }{2\pi }\int
d\phi Yf^{\prime \prime \prime }\,,  \notag \\
\left\{ j[Y],z[h]\right\} &=&z\left[ [Y,h]\right] -\frac{k\alpha_2 }{2\pi }\int
d\phi Yh^{\prime \prime \prime }\,, 	\eqlabel{algebra1} \\
\left\{ p[f_{1}],p[f_{2}]\right\} &=&z\left[ [f_{1},f_{2}]\right] -\frac{%
k\alpha_2 }{2\pi }\int d\phi f_{1}f_{2}^{\prime \prime \prime }\,,  \notag \\
\left\{ p[f],z[h]\right\} &=&\frac{p}{\ell ^{2}}\left[ [f,h]\right] -\frac{%
k\alpha_1 }{2\pi \ell ^{2}}\int d\phi fh^{\prime \prime \prime }\,,  \notag \\
\left\{ z[h_{1}],z[h_{2}]\right\} &=&\frac{z}{\ell ^{2}}\left[ [h_{1},h_{2}]%
\right] -\frac{k\alpha_2 }{2\pi \ell ^{2}}\int d\phi h_{1}h_{2}^{\prime \prime
\prime }\,,  \notag
\end{eqnarray}%
where here $\left[ x,y\right] =xy^{\prime }-yx^{\prime }$ denotes the Lie bracket. By expanding in Fourier modes and defining
\begin{equation*}
\mathcal{J}_{m}=j[e^{im\text{$\phi $}}]\text{\thinspace \thinspace
\thinspace \thinspace ,\thinspace \thinspace \thinspace \thinspace }\mathcal{%
P}_{m}=p[e^{im\text{$\phi $}}]\text{\thinspace \thinspace \thinspace
\thinspace ,\thinspace \thinspace \thinspace \thinspace }\mathcal{Z}%
_{m}=z[e^{im\text{$\phi $}}]\,,
\end{equation*}%
the algebra \eqref*{algebra1} takes the following form:
\begin{equation}
\begin{array}{lcl}
i\left\{ \mathcal{J}_{m},\mathcal{J}_{n}\right\} & = & \left( m-n\right)
\mathcal{J}_{m+n}+\dfrac{c_{1}}{12}\,m^{3}\delta _{m+n,0}\,,\medskip \\[6pt]
i\left\{ \mathcal{J}_{m},\mathcal{P}_{n}\right\} & = & \left( m-n\right)
\mathcal{P}_{m+n}+\dfrac{c_{2}}{12}\,m^{3}\delta _{m+n,0}\,,\medskip \\[6pt]
i\left\{ \mathcal{P}_{m},\mathcal{P}_{n}\right\} & = & \left( m-n\right)
\mathcal{Z}_{m+n}+\dfrac{c_{3}}{12}\,m^{3}\delta _{m+n,0}\,,\medskip \\[6pt]
i\left\{ \mathcal{J}_{m},\mathcal{Z}_{n}\right\} & = & \left( m-n\right)
\mathcal{Z}_{m+n}+\dfrac{c_{3}}{12}\,m^{3}\delta _{m+n,0}\,,\medskip \\[6pt]
i\left\{ \mathcal{P}_{m},\mathcal{Z}_{n}\right\} & = & \medskip \dfrac{1}{%
\ell ^{2}}\left( m-n\right) \mathcal{P}_{m+n}+\dfrac{c_{2}}{12\ell ^{2}}%
\,m^{3}\delta _{m+n,0}\,, \\[6pt]
i\left\{ \mathcal{Z}_{m},\mathcal{Z}_{n}\right\} & = & \dfrac{1}{\ell ^{2}}%
\left( m-n\right) \mathcal{Z}_{m+n}+\dfrac{c_{3}}{12\ell ^{2}}\,m^{3}\delta
_{m+n,0}\,,%
\end{array}
\eqlabel{vir3}
\end{equation}
where the central charges $c_1$, $c_2$ and $c_3$ are related to the CS level $k$ and the invariant tensor constants $\alpha_0$, $\alpha_1$ and $\alpha_2$ defined in \eqref{redef} by
\begin{equation}\eqlabel{ciai}
c_i=12k\alpha_{i-1}\;.
\end{equation}
This structure corresponds to a infinite-dimensional enhancement of the AdS-Lorentz algebra. In the same way as the AdS-Lorentz algebra defines a semi-simple enlargement of the Poincaré symmetry, the algebra \eqref*{vir3} defines a semi-simple enlargement of the $\mathfrak{bms}_3$ symmetry. Such infinite-dimensional symmetry has been
first introduced in~\cite{Caroca:2017onr} as an expansion of the Virasoro algebra.
Here, we have shown that it can be realized as the asymptotic
symmetry of AdS-Lorentz CS gravity theory in three space-time dimensions.
One can see that the AdS-Lorentz algebra is a finite subalgebra spanned by
the generators $\left\{ \mathcal{J}_{0},\mathcal{J}_{1},\mathcal{J}_{-1},%
\mathcal{P}_{0},\mathcal{P}_{1},\mathcal{P}_{-1},\mathcal{Z}_{0},\mathcal{Z}%
_{1},\mathcal{Z}_{-1}\right\} $. It is worth noting that the algebra \eqref*{vir3} is isomorphic to the direct product of three-copies of the Virasoro algebra. In fact, defining new generators
\begin{equation}
\begin{tabular}{lll}
$\mathcal{L}^{+}_{m}=\dfrac{1}{2}\left( \ell ^{2}\mathcal{Z}%
_{m} + \ell \mathcal{P}_{m}\right) \,,$ & $\mathcal{L}^{-}_{m}=\dfrac{1}{2}\left( \ell ^{2}\mathcal{Z}_{-m}-\ell \mathcal{P}_{-m}\right) \,,$ & $\hat{\mathcal{L}}_{m}=%
\mathcal{J}_{-m}-\ell ^{2}\mathcal{Z}_{-m}\,,$%
\end{tabular}%
\end{equation}%
three copies of the Virasoro algebra are revealed:
\begin{equation}\eqlabel{3virasoro}
\begin{array}{lcl}
i\left\{ \mathcal{L}^{+}_{m},\mathcal{L}^{+}_{n}\right\}  & = & \left( m-n\right)
\mathcal{L}^{+}_{m+n}+\dfrac{c^+}{12}m^{3}\delta _{m+n,0}\,, \\[5pt]
i\left\{ \mathcal{L}^{-}_{m},\mathcal{L}^{-}_{n}\right\}  & = & \left(
m-n\right) \mathcal{L}^{-}_{m+n}+\dfrac{c^-}{12}m^{3}\delta _{m+n,0}\,,
\\[5pt]
i\left\{ \hat{\mathcal{L}}_{m},\hat{\mathcal{L}}_{n}\right\}  & = &
\left( m-n\right) \hat{\mathcal{L}}_{m+n}+\dfrac{\hat{c}}{12}m^{3}\delta
_{m+n,0}\,,%
\end{array}%
\end{equation}%
where the central charges are given by
\begin{equation}\eqlabel{cpmandhatc}
c^\pm=\dfrac{1}{2}\left( \ell ^{2}c_{3}\pm\ell c_{2} \right) \;,\quad \hat{c}=\left( c_{1}-\ell ^{2}c_{3}\right) \,.
\end{equation}%
Note that the occurrence of three copies of the Virasoro algebra could have been obtained in a more straightforward way by recalling that the AdS-Lorentz algebra in three dimensions is isomorphic to three copies of the $\mathfrak{so}\left(2,1\right)\simeq \mathfrak{sl}\left(2,\mathbb{R}\right)$ algebra. In fact, using the basis $\{ J^{\pm}_a,\hat{J}_a \} $ defined in \eqref{3lorentz}, the non-vanishing components of the invariant tensor read
\begin{equation}\eqlabel{invten3sl}
\left\langle J_{a}^{\pm}J_{b}^{\pm}\right\rangle =\frac{1}{2}\left(\mu_{2}\pm\mu_{1}\right)\eta_{ab}
\;,\quad
\left\langle \hat{J}_{a}\hat{J}_{b}\right\rangle = \left(\mu_{0}-\mu_{2}\right)\eta_{ab}\;,
\end{equation}
and the connection \eqref*{1f} takes the form $A=A^+ +A^- + \hat{A}$, where
\begin{equation}
A^{\pm}=\left(\omega^{a} \pm \frac{e^{a}}{\ell}+\frac{\sigma^{a}}{\ell^{2}}\right)J^{\pm}_a\;,\quad\hat{A}=\omega^{a}\hat{J}_a\;.
\end{equation}
The action \eqref*{gcs} then splits into the sum of three $SL(2,\mathbb R )$ Chern-Simons actions, one for each $\mathfrak{sl}\left(2,\mathbb{R}\right)$ connection.  They naturally lead to three sets of conserved charges of the form \eqref*{Qcs}, and the boundary conditions for the radial-independent connection \eqref*{aadslor} leads to three $\mathfrak{sl}\left(2,\mathbb{R}\right)$ connections satisfying Brown-Henneaux boundary conditions:
\begin{equation}
\begin{array}{llll}
a^{\pm\,0} & =\mathcal{L}^\pm(x^\pm) dx^{\pm} \,,\quad &  \hat{a}^{0} & = \mathcal{L}(\phi)d\phi,\\
a^{\pm\,1} & =dx^{\pm}\,,\quad &  \hat{a}^{1} & =d\phi\,,\\
a^{\pm\,2} & =0\,,  & \hat{a}^{2} & =0\,.
\end{array}
\end{equation}
where the functions $\mathcal{L}^\pm$ and $\mathcal{L}$ have been defined in \eqref[s]{Lpm} and \eqref*{BMS8}. Therefore, each set of charges coming from \eqref*{Qcs} will satisfy a Virasoro algebra. The corresponding central charges take the standard form: $c^\pm=6k\eta^{ab}\left\langle J_{a}^{\pm}J_{b}^{\pm} \right\rangle\,,\;\hat{c}=6k\eta^{ab}\left\langle \hat{J}_{a}\hat{J}_{b}\right\rangle$. Using the invariant tensor \eqref*{invten3sl} and the definition \eqref*{ciai}, they can be shown to match \eqref{cpmandhatc}, recovering the result \eqref*{3virasoro}.

It is important to remark that, even though the asymptotic symmetry algebra looks simpler in the form \eqref*{3virasoro}, the analysis in the basis $\left\{ J_a,P_a,Z_a\right\}$ is important due to the following reasons: {\it i)} it defines boundary conditions for the fields $e$, $\omega$ and $\sigma$ present in the action \eqref*{cs}, which is the original form of three-dimensional AdS-Lorentz gravity theory that has been previously worked out in the literature (see \cite{Concha:2014zsa,Fierro:2014lka} and references therein); {\it ii)} it allows one to understand the asymptotic symmetry algebra of the theory as a semi-simple enlargement of $\mathfrak{bms}_3$ in the very same way as the Ads-Lorentz algebra has been defined as a semi-simple enlargement of the Poincaré algebra \cite{Soroka:2006aj,Gomis:2009dm,Soroka:2011tc}; {\it iii)} as it will be explained in the following, keeping the analysis in terms of the basis \eqref*{adsl} allows one to make a transparent contact with the deformed $\mathfrak{bms}_3$ algebra obtained in \cite{Concha:2018zeb, Caroca:2017onr}.

\subsection{Flat limit}
		
		The vanishing cosmological constant
		limit $\ell \rightarrow \infty$ can be performed transparently throughout all the steps followed in obtaining the asymptotic symmetry algebra \eqref*{vir3}. In
		particular, this limit applied to connection \eqref*{connection} and their variations \eqref*{var1},\eqref*{var2} lead to the asymptotic form of the Maxwell
		gravity connection introduced in \cite{Concha:2018zeb, Caroca:2017onr}:
		
\begin{equation}
\begin{array}{lcl}
i\left\{ \mathcal{J}_{m},\mathcal{J}_{n}\right\} & = & \left( m-n\right)
\mathcal{J}_{m+n}+\dfrac{c_{1}}{12}\,m^{3}\delta _{m+n,0}\,,\medskip \\[6pt]
i\left\{ \mathcal{J}_{m},\mathcal{P}_{n}\right\} & = & \left( m-n\right)
\mathcal{P}_{m+n}+\dfrac{c_{2}}{12}\,m^{3}\delta _{m+n,0}\,,\medskip \\[6pt]
i\left\{ \mathcal{P}_{m},\mathcal{P}_{n}\right\} & = & \left( m-n\right)
\mathcal{Z}_{m+n}+\dfrac{c_{3}}{12}\,m^{3}\delta _{m+n,0}\,,\medskip \\[6pt]
i\left\{ \mathcal{J}_{m},\mathcal{Z}_{n}\right\} & = & \left( m-n\right)
\mathcal{Z}_{m+n}+\dfrac{c_{3}}{12}\,m^{3}\delta _{m+n,0}\,,\medskip \\[6pt]
i\left\{ \mathcal{P}_{m},\mathcal{Z}_{n}\right\} & = & 0\,, \\[6pt]
i\left\{ \mathcal{Z}_{m},\mathcal{Z}_{n}\right\} & = & 0\,.%
\end{array}
\end{equation}

Note that after such limit, the $\mathfrak{bms}_3$ algebra is recovered by setting the generators $\mathcal{Z}_n$ and the central charge $c_3$ to zero. It is important to remark that this flat limit of AdS-Lorentz CS gravity can be applied only in the $\left\{ J_{a},P_{a},Z_{a}\right\} $ basis.

The relation between the algebra obtained here and the
		deformed $\mathfrak{bms}_{3}$ can be generalized to other algebras. As was discussed in~\cite{Caroca:2017onr}, such infinite-dimensional
		symmetries belong to a family of infinite-dimensional algebras denoted as $%
		\mathfrak{vir}_{\mathfrak{C}_{r}}$ which is related to a generalized $%
		\mathfrak{bms}_{3}$ algebra ($\mathfrak{vir}_{\mathfrak{B}_{r}}$) by an IW
		contraction. \ In particular, $r=4$ corresponds to our result, while $r=3$
		reproduces the 2D-conformal algebra, whose flat limit is precisely given by the $\mathfrak{bms}_{3}$
		algebra. This particular notation is motivated by the fact that they
		correspond to infinite-dimensional lifts of the $\mathfrak{C}_{r}$ and $%
		\mathfrak{B}_{r}$ symmetries, respectively~\cite{Concha:2016hbt}.
		
\newsection{Comments and possible developments}
\label{comments}
In this paper we have studied the asymptotic structure of a CS gravity theory invariant under the semi-simple enlargement of the Poincaré algebra in three-dimensions. In order to carry out the analysis, a generalization of the three-dimensional BMS gauge familiar from the GR analysis~\cite{Barnich:2013yka} has been considered to include the extra field content present in the AdS-Lorentz connection. We have contrasted our results with the stationary solution already known in ADM coordinantes in \cite{Hoseinzadeh:2014bla} and found that it can be recovered as a particular case of our BMS-like extension when a suitable gauge fixing is chosen for the stationary $\bar{\sigma}$ field in \eqref{gauge_fix}. Using this generalized anstaz we have defined boundary conditions for the theory and the CS field equations can be solved exactly, which determines the solution space of the theory completely.
		
The asymptotic symmetry of the theory is given by an infinite dimensional algebra, which defines a semi-simple enlargement of the $\mathfrak{bms}_3$ symmetry in the same spirit as the AdS-Lorentz algebra is a semi-simple enlargement of the $\mathfrak{iso}(2,1)$~\cite{Soroka:2011tc}. This novel infinite dimensional algebra has three central charges and it is isomorphic to three-copies of the Virasoro algebra.
		
The BMS formulation is known for providing a well-defined flat limit at the level of the asymptotic charges when passing from asymptotically AdS to asymptotically flat GR \cite{Barnich:2012aw}. This is also the case in our analysis, where the limit $\ell\rightarrow\infty$ leads to the deformed $\mathfrak{bms}$ algebra previously found in \cite{Concha:2018zeb} as the asymptotic symmetry of a CS theory invariant under the Maxwell algebra. Remarkably, the flat behavior of AdS-Lorentz CS gravity is inherited to its asymptotic symmetry:
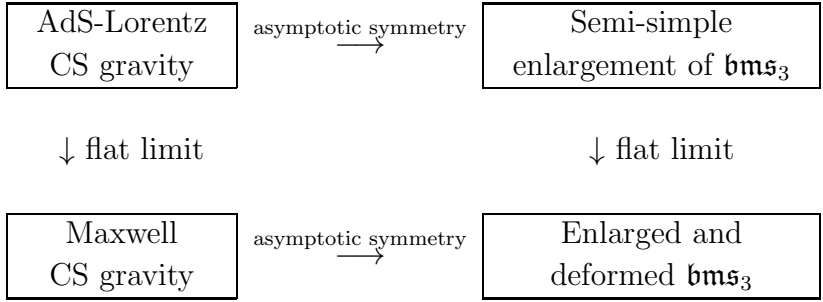
\begin{equation*}
		\begin{tabular}{ccc}
		\cline{1-1}\cline{3-3}
		\multicolumn{1}{|c}{$%
			\begin{array}{c}
			\text{AdS-Lorentz} \\
			\text{CS gravity}%
			\end{array}%
			$} & \multicolumn{1}{|c}{$\overset{\text{asymptotic symmetry}}{%
				\longrightarrow }$} & \multicolumn{1}{|c|}{$%
			\begin{array}{c}
			\text{Semi-simple} \\
			\text{enlargement of }\mathfrak{bms}_{3}%
			\end{array}%
			$} \\ \cline{1-1}\cline{3-3}
		&  &  \\
		$\ \ \downarrow $ $\text{flat limit}$ &  & $\ \ \downarrow $ %
		\text{flat limit} \\
		&  &  \\ \cline{1-1}\cline{3-3}
		\multicolumn{1}{|c}{$%
			\begin{array}{c}
			\text{Maxwell} \\
			\text{CS gravity}%
			\end{array}%
			$} & \multicolumn{1}{|c}{$\overset{\text{asymptotic symmetry}}{%
				\longrightarrow }$} & \multicolumn{1}{|c|}{$%
			\begin{array}{c}
			\text{Enlarged and} \\
			\text{deformed }\mathfrak{bms}_{3}%
			\end{array}%
			$} \\ \cline{1-1}\cline{3-3}
		\end{tabular}%
		\end{equation*} This flat limit was already discussed at the level of the algebras in \cite{Caroca:2017onr}. However, this is the first report showing that the semi-simple enlargement of the $\mathfrak{bms}_3$ algebra (called generalized Virasoro algebra in \cite{Caroca:2017onr}) is the asymptotic symmetry of the a three-dimensional CS gravity theory.
		
The results presented here could be generalized to the case of supersymmetric extensions of AdS-Loretz CS gravity. One would expect to obtain a supersymmetric extension of the asymptotic symmetry algebra found here. Furthermore, one could argue that the flat limit also works at the supersymmetric level. Indeed, it is known that AdS-Lorentz and the Maxwell CS supergravities are related by an Inönü-Wigner contraction ~\cite{Fierro:2014lka,Concha:2016zdb,Concha:2018jxx}.
		
It is important to note that the $\mathfrak{bms}_3$ algebra can also be realized in non-gravitational physics. Indeed, in \cite{Batlle:2017llu} it is shown that the non-centrally extended $\mathfrak{bms}_3$ algebra can be canonically realized as a symmetry of the free Klein-Gordon field in 2 + 1 dimensions. Therefore, as the flat limit of the AdS-Lorentz algebra, the Maxwell algebra, extends the Poincar\'e symmetry to describe particle systems in the presence of constant electromagnetic fields (see for instance \cite{Schrader:1972zd,Bacry1970,Gomis:2017cmt}), it would be interesting to explore the possibility to realize the semi-simple enlargement of $\mathfrak{bms}_3$ here presented, or its flat limit, in scalar field models including the coupling to electromagnetic fields.

\newsection{Acknowledgment}

The authors would like to thank J. Gomis for valuable discussion and comments.  We wish to specially thank O. Miskovic for taking active part in the initial
stages of this project. This work was supported by the Chilean
FONDECYT Grants N$^{\circ}$3160581, N$^{\circ}$3170437, N$^{\circ}$3170438, N$^{\circ}$3160437 and the Grant
VRIIP-UNAP N$^{\circ}$0258-18. P. Salgado-Rebolledo acknowledges DI-VRIEA for financial support through Proyecto Postdoctorado 2018 VRIEA-PUCV.
\bibliographystyle{utphys}
\bibliography{ASM}

\end{document}